\documentclass[%
apl,
 jmp, 
 amsmath,amssymb,
reprint,
tightenlines,
superscriptaddress,
onecolumn,
notitlepage,
11pt
]{revtex4-1}
\usepackage{graphicx}
\usepackage{subfigure}
\usepackage{mathrsfs}
\usepackage[bookmarks=true,colorlinks=true,citecolor=blue,linkcolor=blue,urlcolor=magenta]{hyperref}

\usepackage[english]{babel} 
\usepackage{multirow}
\usepackage{dcolumn}
\usepackage{bm}

\graphicspath{{figures/}}

\usepackage{color}
\usepackage{soul}

\begin{document}

\title[]{The state of nonrelativistic quantum system in a relativistic reference frame}

\author{I.V.~Sharph}
\affiliation{Odessa National Polytechnic University, Shevchenko av. 1, Odessa, 65044, Ukraine.}%

\author{M.A.~Deliyergiyev}%
\affiliation{ Department of Experimental Particle Physics, Jozef Stefan Institute, 
     Jamova 39, SI-1000 Ljubljana, Slovenia.}%

\author{A.G.~Kotanzhyan}%
\affiliation{ Odessa National Polytechnic University, Shevchenko av. 1, Odessa, 65044, Ukraine.}%

\author{K.K.~Merkotan}%
\affiliation{ Odessa National Polytechnic University, Shevchenko av. 1, Odessa, 65044, Ukraine.}%

\author{N.O.~Podolian}%
\affiliation{ Odessa National Polytechnic University, Shevchenko av. 1, Odessa, 65044, Ukraine.}%

\author{O.S.~Potiyenko}%
\affiliation{ Odessa National Polytechnic University, Shevchenko av. 1, Odessa, 65044, Ukraine.}%

\author{D.A.~Ptashynskyy}%
\affiliation{ Odessa National Polytechnic University, Shevchenko av. 1, Odessa, 65044, Ukraine.}%

\author{G.O.~Sokhrannyi}%
\affiliation{ Odessa National Polytechnic University, Shevchenko av. 1, Odessa, 65044, Ukraine.}%

\author{A.V.~Tykhonov}%
\affiliation{ 
D\'{e}partement de physique nucl\'{e}aire et corpusculaire, Universit\'{e} de Gen\'{e}ve, CH-1211 Geneva 4, Switzerland.}%

\author{Yu.V.~Volkotrub}%
\affiliation{ Odessa National Polytechnic University, Shevchenko av. 1, Odessa, 65044, Ukraine.}%

\author{V.D.~Rusov}%
 \email{siiis@te.net.ua}
\affiliation{Odessa National Polytechnic University, Shevchenko av. 1, Odessa, 65044, Ukraine.}%
\affiliation{Department of Mathematics, Bielefeld University, 
      Universitatsstrasse 25, 33615 Bielefeld, Germany.}%



\begin{abstract}
We consider the problem of internal particle state transformation, which is a bound state of several constituents, from the particle's rest frame to the system in which this particle is relativistic. It is assumed that in the rest frame of the composite particle, its internal state can be considered in the nonrelativistic approximation. It is shown, that this internal state is unchanged during the transition from one reference frame to another. Namely, given the particle is spherically symmetric in the rest frame, it remains spherically symmetric in any other reference frame, and does not undergo Lorentz contraction along the direction of motion of moving reference frame with respect to the rest frame. We discuss a possible application of these results to the description of hadron-hadron scattering, considering hadrons as a bound states of quarks.
\end{abstract}

\maketitle

%


\section{Introduction}
Recently it has been shown that the hadron-hadron inelastic scattering processes can be successfully described with the Laplace's method~\cite{ujp}. However, calculations in~\cite{ujp} have been performed using a simplified scalar theory, rather than the quantum chromodynamics (QCD). This allowed to reproduce the experimental data for hadron-hadron scattering cross section on a qualitative level only~\cite{cej}. Obtained results force us to try to apply this method within the framework of more realistic theory, in QCD. Herewith we run into the well-known problem: there are quark and gluon lines in Feynman diagrams in QCD,  whereas there are no lines representing hadrons. 

The problem of relativistic description of hadrons as a bound states of quarks and gluons has been scrutinized for a long time \cite{Brodsky:1997de}. It is related primarily to the nature of relativistic many-particle state, hence, to the necessity of identifying a large number of interconnected probability amplitudes, which describe the Fock state of the system \cite{Brodsky:1997de, Strikman:2011zz, Strikman:2011cx, Diehl:2011yj}. 
However, given there is a free hadron in its rest frame in the initial (final) state of scattering process, we can try to simplify the problem using a non-relativistic approximation. That is, we consider hadron as a non-relativistic compound state of a certain number of constituent quarks of certain flavors.
Still, initial (final) state of the scattering process comprises more than one hadron. Therefore, it is impossible to pick up the reference frame which would be the rest frame for all the hadrons under investigation simultaneously. As a result, the problem arises of transforming the nonrelativistic internal state and Hamiltonian of particle, as we pass from its rest frame to the reference frame where this particle is relativistic.
The essence of this problem can be illustrated with the following simple example. Consider the most trivial quantum system $\textendash$ the hydrogen atom in the simplest spherically symmetric ground state. Imagine an inertial observer moving with relativistic velocity with respect to it. The observer measures the coordinates (momenta) of particles that make up the system. What would be the result of this measurement?
In other words, how does the probability amplitude describing these measurements look like, and how does this probability amplitude depend on the probability amplitude in the rest frame (in a center-of-mass system of particles comprising the atom)? 
Further in the paper, for instance, will be considered a meson as a two-particle system of quark and antiquark, and afterwards we will apply obtained results to a more complex three-quark systems, i.e. baryons.

Outlined problem is rather non-typical. Usually one deals with the measurements of a specific quantities associated with the same event, which are performed in the different reference frames. This is not the case if we deal with the probability amplitude for many-particle system. Consider two inertial observers, which we call \textit{unprimed} and \textit{primed} respectively. Then, the probability amplitude for two-particle system in the reference frame of \textit{unprimed} observer (denoted as $\Psi \left( t,{{{\mathbf{r}}}_{1}},{{{\mathbf{r}}}_{2}} \right)$) describes the result of simultaneous measurement of particles' coordinates, performed at a certain moment of time $t$ in this frame. 
Likewise, the probability amplitude ${\Psi }'\left( {t}',{{{{\mathbf{r}}'}}_{1}},{{{{\mathbf{r}}'}}_{2}} \right)$ in the reference frame of the \textit{primed} observer describes the results of measurements, which are simultaneous with respect to this observer at his clock time ${t}'$. However, measurements that are simultaneous in the reference frame of the \textit{primed} observer will not be simultaneous in the reference frame of the \textit{unprimed} one, and vice versa. 
In this way, the considered problem essentially differs from the classical problem of Lorentz contraction. 
In case of Lorentz contraction, measurements of coordinates of rod's ends must be synchronized in the reference frame of moving observer, whereas the corresponding measurements in the rod's rest frame are not necessarily synchronized, whereby rod's length is calculated through the same pair of events in two different frames. 
In our case, the probability amplitudes $\Psi \left( t,{{{\mathbf{r}}}_{1}},{{{\mathbf{r}}}_{2}} \right)$ and ${\Psi }'\left( {t}',{{{{\mathbf{r}}'}}_{1}},{{{{\mathbf{r}}'}}_{2}} \right)$ are associated with the different realizations of the measurement process. 
As a result, one cannot conclude any kind of relation between the magnitudes $t$ and ${t}'$, since such a relations could be established only between the time coordinates of the same event, measured with respect to different reference frames. Accordingly, it is impossible to establish a tie between the values of ${{\mathbf{r}}_{1}}$ and ${{\mathbf{r}}_{2}}$, as well as between ${{{\mathbf{r}}'}_{1}}$ and ${{{\mathbf{r}}'}_{2}}$. 
Hence, there is no relations similar to Lorentz transformations between the arguments of the probability amplitudes in both reference frames. 
Therefore, the notions of length contraction and time dilation, which are the consequences of Lorentz transformations, are not applicable in our case. 
Note also, that the inclusion of ``timelike" coordinates, like in quantization on a light-cone \cite{Brodsky:1997de}, as well as dealing with the theory on arbitrary spacelike surface, which is done in quasi-potential approach \cite{MatSavSis02}, does not resolve the aforementioned problem of simultaneity, since both are based on the assumption that the relation exists between the arguments of probability amplitudes in different reference frames.

Thus, in the problem of state transformation at the transition from one inertial reference frame to another, it is improper to consider connections between the values of probability amplitudes, corresponding to the same events in different reference frames. Rather, one should examine the relation between the values of probability amplitude in the different reference frames, corresponding to the same values of arguments, similar to that in dealing with the internal symmetries. Taking this into account, we will denote the probability amplitude in the \textit{primed} reference frame through ${\Psi }'\left( t,{{{\mathbf{r}}}_{1}},{{{\mathbf{r}}}_{2}} \right)$. 

The principle method to solve the state transformation problem at the transition from one inertial reference frame to another is provided by the field quantization postulate, established in \cite{Bogolubov}. According to this postulate, generators of transformation of a Fock space elements at the transition from one reference frame to another, where transition consists of boost and rotations, correspond  to the components of the angular momentum tensor of the system of fields that make up the state of the system.  In particular, if $\left| \Psi \right\rangle$ denotes the Fock state in some inertial reference frame, and $\left| \Psi^{\prime} \right\rangle$ denotes the same state, but in another reference frame, which can be obtained from the initial one through the boost transformation along the $z$-axis with rapidity $Y$, then, according to the postulate \cite{Bogolubov}, we have:
\begin{equation}
\left| \Psi^{\prime} \right\rangle =\exp \left( i{{{\hat{M}}}_{03}}Y \right) \left| \Psi \right\rangle,
\label{eq:Fokstan}
\end{equation}
where $\hat{M}_{03}$ is the operator of $z$-component of angular momentum tensor, which can be represented as integral of the corresponding density $\hat{M}_{03}^{0}\left( t,{{{\mathbf{r}}}} \right)$ in a three-dimensional coordinate space:
\begin{equation}
\hat{M}_{03} = \int{\hat{M}_{03}^{0}\left( t,{{{\mathbf{r}}}} \right)  d\mathbf{r}}
\label{eq:OperatorAngMomentumTensor}
\end{equation}
If we consider relations Eq.\ref{eq:Fokstan} and Eq.\ref{eq:OperatorAngMomentumTensor} in the Heisenberg representation, then both states $\left| \Psi \right\rangle$ and $\left| \Psi^{\prime} \right\rangle$ are time-independent due to reciprocal representation. 
Likewise, in the Heisenberg picture $\hat{M}_{03}$ is time-independent as well, since, according to the Noether's theorem, this quantity is the integral of motion. 
The density function $\hat{M}_{03}^{0}\left( t,{{{\mathbf{r}}}} \right)$ is the only quantity which depends on time. 
As seen from  Eq.\ref{eq:OperatorAngMomentumTensor}, the values of density function under the integration sign are taken for different points in space, but in the same moment of time in the reference frame, where the density function is defined.

Next, if we apply the tensor-transformation rules to the components of the angular momentum tensor $\hat{M}_{ab}$ at the transition from the reference frame, with respect to which $\left| \Psi \right\rangle$ is defined, to the one, where $\left| \Psi^{\prime} \right\rangle$ is defined, we get:
\begin{equation}
\hat{M}_{03}^{\prime} = \hat{M}_{03}
\label{eq:TransitionBetweenRefFrames}
\end{equation}
That is, in Eq.\ref{eq:Fokstan} we can equally use either $\hat{M}_{03}$ as a generator written with respect to the original reference frame, or as generator written with respect to the new reference frame. 
One may note however, that if we write generator $\hat{M}_{03}^{\prime}$ as an integral of the corresponding density in the new reference frame, then we have to consider only those values of density, that are synchronous with respect to this reference frame. 
Again, in this case we encounter the same problem: with the inability to ensure the simultaneity with respect to two different reference frames. That issue force us to consider the density function in the original reference frame at the whole space-time domain, rather than in a subset of this domain, over which integration in Eq.\ref{eq:OperatorAngMomentumTensor} is done. 
Afterwards, according to the tensor transformation rules, this density function should be transformed into the new reference frame. 
Part of this transformation pertains to the expression of arguments of density function through the space-time coordinates in the new reference frame. 
Thereafter, it is necessary to separate out a subset of space-time domain, comprising all the points that have the same time coordinate in the new reference frame, and carry out integration over this subset, obtaining thereby the generator $\hat{M}_{03}^{\prime}$. From these reflections, it is clear that in the different reference frames the integration of density function is carried out over the different subsets of space-time. Therefore, it is impossible to interconnect the arguments of density function in the different reference frames via the Lorentz transformations, as it would be the cases if one integrates over a single set, which is described using coordinates of different reference frames.

On the other hand, the density $\hat{M}_{03}^{\prime~0}\left( t^{\prime},{{{\mathbf{r^{\prime}}}}} \right)$ can be obtained directly with the help of the Noether's theorem from Lagrangian and field equations written in the new reference frame. 
Given the fact that all relations have the same form in all inertial reference frames, we can write $\hat{M}_{03}^{0}$ with respect to original reference frame, and $\hat{M}_{03}^{\prime~0}$ generator written with respect to a new reference frame. 
From principle of relativity functions $\hat{M}_{03}^{0}\left( t,{{{\mathbf{r}}}} \right)$  and $\hat{M}_{03}^{\prime~0}\left( t^{\prime},{{{\mathbf{r^{\prime}}}}} \right)$ are the same function, but each of its own variables. 
Now, let us recall that it is impossible to establish any relation between $t$ and $t^{\prime}$, as well as the fact that the result of the integration of the angular-momentum's density does not depend on the exact choice of time moment when the integration is done, in both reference frames. 
With this in mind, we can set $t^{\prime} = t$. 
Then it imminently follows, that two quantities in Eq.\ref{eq:OperatorAngMomentumTensor}, each corresponds to its own reference frame, will differ only in the notation of integration variables. 
This in turn means that in Eq.\ref{eq:Fokstan} the components of $\left| \Psi \right\rangle$ and $\left| \Psi^{\prime} \right\rangle$ Fock states are taken with the same values of arguments, however these values define the quantities, that defined with respect to the different reference frames.

Transformation of states under rotations does not interfere with the described problems. 
Therefore, to solve the problem 
of state transformation upon arbitrary Lorentz transformation, it is enough to consider it for the boost. 
In order to simplify this problem of application of a nonrelativistic approximation we must construct the appropriate nonrelativistic approximation for the $\hat{M}_{03}$ generator. 

Different ways of formulating such an approximation we will consider in the further sections. 
But even after the transition to the nonrelativistic approximation, still the the problem remains of acting with the exponent in operator from Eq.\ref{eq:Fokstan} on state of the bound system in original reference frame. 
To make it easier, let us make the following considerations.

Let us consider the hadron in its rest frame. In this system the state $\left| \Psi \right\rangle$ must be eigenstate for total momentum $\hat{\mathbf{P}}$ of the all particles that make up this system. And at the same time this state must correspond to zero eigenvalue. 
Before the transition to non-relativistic approximation the temporal progress of state $\left| \Psi \right\rangle$ of the particles 
system, that make up the hadron, can be represented in next form
\begin{equation}
\left| \Psi \left( t \right) \right\rangle =\exp \left( -i\hat{H}t \right)\left| \Psi \left( t=0 \right) \right\rangle ,
\end{equation}
where $\hat{H}$ is the relativistic Hamiltonian of a system of fields, whose quants are making up the hadron. In the reference frame that obtained from outcome system by applying the boost, according to \cite{Bogolubov} we will obtain:
\begin{equation}
\left| {\Psi }'\left( t \right) \right\rangle =\hat{U}\left( Y \right)\left( \exp \left( -i\hat{H}t \right)\left| \Psi \left( t=0 \right) \right\rangle  \right).
\label{eq:peretvorenna}
\end{equation}
Here $\hat{U}\left( Y \right)$ is the unitary state transformation operator of Eq.\ref{eq:Fokstan} from \cite{Bogolubov} as a result of boost with rapidity $Y$. 
\begin{align}
\hat{U}\left( Y \right)=\exp \left( i{{{\hat{M}}}_{03}}Y \right).
\label{eq:generator}
\end{align}

Taking into account that we are dealing with the eigenstate of total momentum of the system and this state corresponds the zero eigenvalue we can write relation Eq.\ref{eq:peretvorenna} in the following form:
\begin{eqnarray}
& \left| {\Psi }'\left( t \right) \right\rangle =\hat{U}\left( Y \right)\left( \exp \left( -i\left( \hat{H}t-\left( \hat{\mathbf{P}}\cdot \hat{\mathbf{R}} \right) \right) \right)\left| \Psi \left( t=0 \right) \right\rangle  \right),
\label{eq:peretvorenna1}	
\end{eqnarray}
where $\mathbf{R}$ is a set of the three arbitrary coordinates. The specific choice of these coordinates is not sufficient, because the eigenvalue of total momentum operator $\hat{\mathbf{P}}$ is zero.

Now we rewrite expression Eq.\ref{eq:peretvorenna1} in another form:
\begin{eqnarray}
& \left| {\Psi }'\left( t \right) \right\rangle =\hat{U}\left( Y \right)\hat{u}\left( x \right){{{\hat{U}}}^{-1}}\left( Y \right)\hat{U}\left( Y \right)\left| \Psi \left( t=0 \right) \right\rangle,
\label{eq:peretvorenna2}	
\end{eqnarray}
where we use the following notations:
\begin{eqnarray}
x\equiv \left( t,{{R}_{x}},{{R}_{y}},{{R}_{z}} \right), ~~ \hat{u}\left( x \right)\equiv \exp \left( -i\left( \hat{H}t-\left( \hat{\mathbf{P}}\cdot \hat{\mathbf{R}} \right) \right) \right).
\label{eq:poznachenna_x_u}	
\end{eqnarray}

Expression $\hat{U}\left( Y \right)\hat{u}\left( x \right){{\hat{U}}^{-1}}\left( Y \right)$ in Eq.\ref{eq:peretvorenna2} formally coincides with the expression which appears at transformation of operator field functions \cite{Bogolubov}. Therefore, denoting the matrix of boost along $z$ direction by ${{\Lambda }^{\left( 0 \right)}}\left( Y \right)$ we get 
\begin{eqnarray}
\hat{U}\left( Y \right)\hat{u}\left( x \right){{\hat{U}}^{-1}}\left( Y \right)=\hat{u}\left( {{\Lambda }^{\left( 0 \right)}}\left( Y \right)x \right).
\label{eq:poznachenna_u1}	
\end{eqnarray}

Then, instead of Eq.\ref{eq:peretvorenna2}, we can write:
\begin{eqnarray}
& \left| {\Psi }'\left( t \right) \right\rangle =\exp \left( -it\left( \operatorname{ch}\left( Y \right)\hat{H}+\operatorname{sh}\left( Y \right){{{\hat{P}}}_{z}} \right) \right)\exp \left( i{{R}_{z}}\left( \operatorname{sh}\left( Y \right)\hat{H}+\operatorname{ch}\left( Y \right){{{\hat{P}}}_{z}} \right) \right)\times\nonumber  \\
& \times \exp \left( i\left( {{R}_{x}}{{{\hat{P}}}_{x}}+{{R}_{y}}{{{\hat{P}}}_{y}} \right) \right)\hat{U}\left( Y \right)\left| \Psi \left( t=0 \right) \right\rangle .
\label{eq:peretvorenna3}	
\end{eqnarray}

Pay attention, that operators $\hat{H}$ and $\hat{\mathbf{P}}$ are included in relation Eq.\ref{eq:peretvorenna3} that belongs to the original reference frame, in which we can apply the nonrelativistic approximation. 
Using this approximation these operators may be replaced by nonrelativistic internal Hamiltonian of the quarks system which make up hadron, and nonrelativistic momentum operator of this system  accordingly. The quantity $\left| \Psi \left( t=0 \right) \right\rangle $ in such nonrelativistic approximation may be replaced by coordinate part of the probability amplitude of energy eigenstate for two-particle system (quark and antiquark). 
In addition, if we consider a extreme case of small rapidities $Y$, one may note that we must choose a coordinates of the center of mass as an arbitrary coordinates of the vector $\hat{\mathbf{R}}$:
\begin{align}
\mathbf{R}=\mathbf{R}\left( {{{\mathbf{r}}}_{1}},{{{\mathbf{r}}}_{2}} \right)=\frac{{{m}_{1}}{{{\mathbf{r}}}_{1}}+{{m}_{2}}{{{\mathbf{r}}}_{2}}}{{{m}_{1}}+{{m}_{2}}}.
\label{eq:centr_mas}
\end{align}

Thus, the simplification discussed above and achieved throughout the Eq.\ref{eq:peretvorenna1}-\ref{eq:peretvorenna3} means that we don't need to describe the transformation of the whole probability amplitude in the energy eigenstate during the transition from the center-of-mass (quark and antiquark) system to another reference system, but we can limit that calculation only to the transformation of the coordinate part of this probability amplitude.

Finally, note that we are interested in the bound systems of particles where the major role is played by the strong interactions. If one will write a component $\hat{M}_{03}$ for QCD Lagrangian, you will notice that nonzero contribution to the spin part of this component is made only by gluon fields. But there is always an option to choose the calibration of gluon fields, in which spin contribution is zero. In particular, in the Hamiltonian calibration scheme \cite{slavnov1991gauge2435572} the zero component of the 4-vector gluon fields is assigned to zero. Therefore, further in work using the $\hat{M}_{03}$ - component of angular momentum, we will consider only the contribution of the orbital moment.

\section {Approximation of Lorentz transformation generators using differential operators}

According to \cite{Bogolubov} the representation of ${{\hat{M}}_{03}}$ by the differential operators looks as follows:
\begin{align}
{{\hat{M}}_{03}}=i\left( t\frac{\partial }{\partial z}+z\frac{\partial }{\partial t} \right).
\label{eq:dif1particl}
\end{align}
Therefore, we note that the representation of generators through differential operators can be obtained by considering some function of coordinates and time with a corresponding substitution of the independent variables in the function. But, as was mentioned in the previous section, in our case the replacement of the independent variables is impossible. Therefore, relation Eq.\ref{eq:dif1particl} can be understood only as a limit to which the ``proper" relativistic operator ${{\hat{M}}_{03}}$ is approached in the transition to a nonrelativistic approximation. The question then arises: ``To what limit should this operator approach in case of many-particle system?". Given that the spatial components of the momentum are presented as a sum of the corresponding single-particle operators, we can make an assumption that the components in which one of indices is equal to zero are also additive. Then, for the two-particle system we have:
\begin{align}
{{\hat{M}}_{03}}=i\left( t\left( \frac{\partial }{\partial {{z}_{1}}}+\frac{\partial }{\partial {{z}_{2}}} \right)+\left( {{z}_{1}}+{{z}_{2}} \right)\frac{\partial }{\partial t} \right).
\label{eq:pripucenna}
\end{align}
As already noted, a quantity $\left| \Psi \left( t=0 \right) \right\rangle $, which is included in Eq.\ref{eq:peretvorenna3}, can be replaced in our case by coordinates part of energy eigenstate, which we denote as $\psi \left( {{{\mathbf{r}}}_{1}},{{{\mathbf{r}}}_{2}} \right)$, at the transition to a nonrelativistic approximation. This function does not depend on time and is the operator eigenfunction of the total momentum of system, that corresponds to zero eigenvalue. If we consider that operator Eq.\ref{eq:pripucenna} can be written as:
\begin{align}
{\hat M_{03}} =  - t{\hat P_z} + \left( {{z_1} + {z_2}} \right)i\frac{\partial }{{\partial t}},
\label{eq:pripucenna1}
\end{align}
Hence, we will come to the conclusion that function $\psi \left( {{{\mathbf{r}}}_{1}},{{{\mathbf{r}}}_{2}} \right)$ is also the eigenfunction of operator ${{\hat{M}}_{03}}$, which corresponds to zero eigenvalue.

This can be explained by the following reflections. Since the original reference frame is the center of mass frame of quark and antiquark, we have: 
\begin{align}
\psi \left( {{{\mathbf{r}}}_{1}},{{{\mathbf{r}}}_{2}} \right)=\psi \left( {{{\mathbf{r}}}_{2}}-{{{\mathbf{r}}}_{1}} \right).
\label{eq:cm_koord_chast}
\end{align}
If in the expression 
\begin{align}
i\left( t\left( \frac{\partial }{\partial {{z}_{1}}}+\frac{\partial }{\partial {{z}_{2}}} \right)+\left( {{z}_{1}}+{{z}_{2}} \right)\frac{\partial }{\partial t} \right)\psi \left( {{{\mathbf{r}}}_{2}}-{{{\mathbf{r}}}_{1}} \right),
\label{eq:M03psi}
\end{align}
from ${{\mathbf{r}}_{1}}$ and ${{\mathbf{r}}_{2}}$ we move to the new variables
\begin{align}
{\mathbf r_ + } = {\mathbf r_1} + {\mathbf r_2},~~~{\mathbf r_ - } = {\mathbf{r_1}} - {\mathbf{r_2}},
\label{eq:novi_rplus_rminus}
\end{align}
in this way, operator in Eq.\ref{eq:M03psi} will depend only on $z$ component of vector $\mathbf{r_ {+}}$ as ${{z}_{+}}$, and function on which this operator acts will depends only on $z$ component of vector $\mathbf{r_ {-}}$ as ${{z}_{-}}$.

Thereby we can make the next conclusion
\begin{align}
 \exp \left( i{{{\hat{M}}}_{03}}Y \right)\psi \left( {{{\mathbf{r}}}_{2}}-{{{\mathbf{r}}}_{1}} \right)=\psi \left( {{{\mathbf{r}}}_{2}}-{{{\mathbf{r}}}_{1}} \right).
\label{eq:visnovok}
\end{align}
Namely, the internal state of meson does not change in the transition to a new reference frame.

In the case of baryons, taking assumption that all components of momentum are additive, instead of Eq.\ref{eq:pripucenna} we obtain:
\begin{equation}
{{{\hat{M}}}_{03}}=i\left( t\left( \frac{\partial }{\partial {{z}_{1}}}+\frac{\partial }{\partial {{z}_{2}}}+\frac{\partial }{\partial {{z}_{3}}} \right) +\left( {{z}_{1}}+{{z}_{2}}+{{z}_{3}} \right)\frac{\partial }{\partial t} \right).
\label{eq:pripucenna3}
\end{equation}
This operator is expressed through the operator of $z$-component of total momentum of the system. Therefore, when we act by this operator on the eigenfunction of the total moment momentum operator that corresponds to zero eigenvalue we get zero.

The considerations in this section suffer two essential shortcomings. First, the ``proper" relativistic operator ${{\hat{M}}_{03}}$ is not realized through differential operators, but it is given in the second quantization representation. Therefore, it is more convenient to look for the nonrelativistic limit of this operator in this representation. In addition, we have essentially used the assumptions Eq.\ref{eq:pripucenna} and Eq.\ref{eq:pripucenna3}. 
These assumptions are not required in the second quantization representation, owing to the fact that the expression for the operators does not depend on whether these operators are set on the single-particle or on the many-particle space. 

Therefore, the arguments in this section may be considered only as auxiliary. In the next section we will demonstrate that considering this problem in the secondary quantization representation one can get the same result.

\section{Approximation of Lorentz transformation generators in the second quantization representation}

We denote ${{\hat{q}}^{+}}\left( f,\nu ,c,\mathbf{r} \right)$ as nonrelativistic quark creation operator in the coordinate representation of the second quantization. Indices $f,\nu,c$ set flavor, spin and color state of quarks, respectively, where quarks are created in the radius eigenvector state corresponding to the eigenvalue $\mathbf{r}$. Creation antiquark operator in the same state denoted as ${{\hat{\bar{q}}}^{+}}\left( f,\nu ,c,\mathbf{r} \right)$ and annihilation operators as ${{\hat{q}}^{-}}\left( f,\nu ,c,\mathbf{r} \right)$ and ${{\hat{\bar{q}}}^{-}}\left( f,\nu ,c,\mathbf{r} \right)$ respectively.

In these notations, the coordinate of the internal state of meson as quark-antiquark system can be represented in the form:
\begin{equation}
\begin{split}
\left| \mu  \right\rangle =\int{d{{{\mathbf{r}}}_{2}}d{{{\mathbf{r}}}_{1}}\psi \left( \left| {{{\mathbf{r}}}_{2}}-{{{\mathbf{r}}}_{1}} \right| \right)}s\left( {{\nu }_{1}},{{\nu }_{2}} \right)c\left( {{c}_{1}},{{c}_{2}} \right)a\left( {{f}_{1}},{{f}_{2}} \right)\times  \\
\times {{{\hat{q}}}^{+}}\left( {{f}_{1}},{{\nu }_{1}},{{c}_{1}},{{{\mathbf{r}}}_{1}} \right){{{\hat{\bar{q}}}}^{+}}\left( {{f}_{2}},{{\nu }_{2}},{{c}_{2}},{{{\mathbf{r}}}_{2}} \right)\left| 0 \right\rangle .
\end{split}
\label{eq:coordinatna_castina}
\end{equation}
In this relation have denoted the spin, color and flavor of probability amplitudes through $s\left( {{\nu }_{1}},{{\nu }_{2}} \right)c\left( {{c}_{1}},{{c}_{2}} \right)a\left( {{f}_{1}},{{f}_{2}} \right)$, respectively, whereas the function $\psi \left( \left| {{{\mathbf{r}}}_{2}}-{{{\mathbf{r}}}_{1}} \right| \right)$ describes the coordinate dependence of probability amplitude in the center of mass frame of quark and antiquark. 
Since we consider the coordinate part of the energy eigenstate as $\psi \left( \left| {{{\mathbf{r}}}_{2}}-{{{\mathbf{r}}}_{1}} \right| \right)$, we should consider the eigenfunctions of the nonrelativistic Hamiltonian of quark and antiquark system. Besides, as is usually assumed the summation goes over repeated indices. Also was used the usual notation for vacuum state $\left| 0 \right\rangle$. 

Since the dependence of all quantities on internal indices is insufficient, we hereinafter denote the set of indices $\left\{ \nu ,c,f \right\}$ by $\xi$, and the dependence of probability amplitude on internal indices as: 
\begin{equation}
s\left( {{\nu }_{1}},{{\nu }_{2}} \right)c\left( {{c}_{1}},{{c}_{2}} \right)a\left( {{f}_{1}},{{f}_{2}} \right)\equiv F\left( {{\xi }_{1}},{{\xi }_{2}} \right).
\label{eq:poznachennaF}	
\end{equation}
Notably, instead of Eq.\ref{eq:coordinatna_castina} we can write
\begin{equation}
\begin{split}
 \left| \mu  \right\rangle =F\left( {{\xi }_{1}},{{\xi }_{2}} \right)\int{d{{{\mathbf{r}}}_{2}}d{{{\mathbf{r}}}_{1}}\psi \left( \left| {{{\mathbf{r}}}_{2}}-{{{\mathbf{r}}}_{1}} \right| \right)}{{{\hat{q}}}^{+}}\left( {{\xi }_{1}},{{{\mathbf{r}}}_{1}} \right){{{\hat{\bar{q}}}}^{+}}\left( {{\xi }_{2}},{{{\mathbf{r}}}_{2}} \right)\left| 0 \right\rangle.
\end{split}
\label{eq:coord_cast}
\end{equation}
As it is known, in the field theory the operator ${{\hat{M}}_{03}}$ is presented in the form: 
\begin{equation}
{{\hat{M}}_{03}}=\int{d\mathbf{r}}\left( {{x}_{3}}{{{\hat{T}}}_{00}}\left( {\mathbf{r}} \right)-{{x}_{0}}{{{\hat{T}}}_{30}}\left( {\mathbf{r}} \right) \right),
\label{eq:vt_kv_generator}	
\end{equation}
where ${{\hat{T}}_{00}}\left( {\mathbf{r}} \right)$ and ${{\hat{T}}_{30}}\left( {\mathbf{r}} \right)$ are the operators of the corresponding components of the energy-momentum tensor, ${{x}_{0}}\equiv t$ is the time component of the coordinate 4-vector, and ${{x}_{3}}\equiv \left( -z \right)$ $\textendash$ is covariant component of the coordinate 4-vector along the $z$-axis. 
Hence, the relation Eq.\ref{eq:vt_kv_generator} clearly can be obviously rewritten in the form:
\begin{equation}
{{\hat{M}}_{03}}=-t{{\hat{P}}_{z}}+\int{d\mathbf{r}}\left( {{x}_{3}}{{{\hat{T}}}_{00}}\left( {\mathbf{r}} \right) \right),
\label{eq:M03_t}	
\end{equation}
where ${{\hat{P}}_{z}}$ is operator of the $z$ component of the total momentum of the system.

Note, that relations Eq.\ref{eq:vt_kv_generator} and Eq.\ref{eq:M03_t} are accurate and do not require any assumptions and approximations. Herewith, the dependence on $t$ in Eq.\ref{eq:M03_t} coincides with Eq.\ref{eq:pripucenna1}, while Eq.\ref{eq:pripucenna1} is a consequence of the assumptions Eq.\ref{eq:pripucenna} and Eq.\ref{eq:pripucenna3}. Thus, we can conclude that Eq.\ref{eq:M03_t} proves the validity of these assumptions.

The state Eq.\ref{eq:coord_cast} is an eigenstate for the total momentum of the system, which corresponds to the zero eigenvalue, so the action of the first summand of Eq.\ref{eq:M03_t} on this state trivially gives zero. Therefore, we will represent the second summand of Eq.\ref{eq:M03_t} as follows 
\begin{equation}
{{\hat{M}}_{03}}\left( {{{\hat{T}}}_{00}} \right)=\int{{{x}_{3}}}{{\hat{T}}_{00}}\left( {\mathbf{r}} \right)d\mathbf{r}.
\label{eq:M03_ot_T00}	
\end{equation}
In order to act by this operator on the state of two-particle system Eq.\ref{eq:coord_cast}, we shall construct a nonrelativistic approximation for the energy density ${{T}_{00}}\left( {\mathbf{r}} \right)$. For the solution of this problem it is most convenient to use the representation of second quantization, because in this representation the Hamiltonian is written as an integral from some operator-valued function, which can be taken as nonrelativistic limit of the energy density.

Nonrelativistic Hamiltonian of the quark-antiquark system in the second quantization representation can be written in the form:
\begin{equation}
\begin{split}
& \hat{H}={{{\hat{H}}}^{\left( 0 \right)}}+{{{\hat{H}}}^{\left( V \right)}}, \\
& {{{\hat{H}}}^{\left( 0 \right)}}=\int{d\mathbf{r}}\left( {{{\hat{\bar{q}}}}^{+}}\left( \xi ,\mathbf{r} \right)\left( -\frac{1}{2m}\Delta  \right){{{\hat{q}}}^{-}}\left( \xi ,\mathbf{r} \right) \right)+\int{d\mathbf{r}}\left( {{{\hat{q}}}^{+}}\left( \xi ,\vec{r} \right)\left( -\frac{1}{2m}\Delta  \right){{{\hat{\bar{q}}}}^{-}}\left( \xi ,\mathbf{r} \right) \right), \\
& {{{\hat{H}}}^{\left( V \right)}}=\frac{1}{2}\int{d{{{\mathbf{r}}}_{1}}d{{{\mathbf{r}}}_{2}}}V\left( {{{\mathbf{r}}}_{2}}-{{{\mathbf{r}}}_{1}} \right){{{\hat{\bar{q}}}}^{+}}\left( {{\xi }_{1}},{{{\mathbf{r}}}_{1}} \right){{{\hat{q}}}^{+}}\left( {{\xi }_{2}},{{{\mathbf{r}}}_{2}} \right){{{\hat{\bar{q}}}}^{-}}\left( {{\xi }_{2}},{{{\mathbf{r}}}_{2}} \right){{{\hat{q}}}^{-}}\left( {{\xi }_{1}},{{{\mathbf{r}}}_{1}} \right),\\
\end{split}
\label{eq:H}
\end{equation}
where $V\left({{{\mathbf{r}}}_{2}}-{{{\mathbf{r}}}_{1}}\right)$ is the potential energy of the quark-antiquark interaction, $m$ is the mass of quark or antiquark, which is approximately independent of the flavor, because the bound state exists due to the strong interaction and other types of interactions are neglected.

As is well known, in the representation of the two-particle Hamiltonian through the differential operators is considered by introducing Jacobi coordinates
\begin{equation}
\mathbf{R}=\frac{1}{2}\left( {{{\mathbf{r}}}_{1}}+{{{\mathbf{r}}}_{2}} \right),~~~\mathbf{r}={{\mathbf{r}}_{2}}-{{\mathbf{r}}_{1}},
\label{eq:Jacobi}	
\end{equation}
which helps to represent the Hamiltonian as a sum of two commuting operators $\textendash$ center of mass Hamiltonian, which depends only on $\mathbf{R}$ and internal Hamiltonian which depends only on $\mathbf{r}$. We want to achieve a similar representation in case when the Hamiltonian is written through the creation and annihilation operators. For this purpose, it is convenient to rewrite one-particle part of Hamiltonian ${{\hat{H}}^{\left( 0 \right)}}$ in the two-particle operator form. As we examine the nonrelativistic approximation, all operators can be considered on the subspace of the Fock space with a fixed number and composition of particles. In our case, we consider the subspace of states which contain one quark and one antiquark. The basis states of this subspace can be written in the form:
\begin{equation}
\left| {{\xi }_{1}},{{\xi }_{2}},{{{\mathbf{r}}}_{1}},{{{\mathbf{r}}}_{2}} \right\rangle ={{\hat{\bar{q}}}^{+}}\left( {{\xi }_{1}},{{{\mathbf{r}}}_{1}} \right){{\hat{q}}^{+}}\left( {{\xi }_{2}},{{{\mathbf{r}}}_{2}} \right)\left| 0 \right\rangle.
\label{eq:bazis}	
\end{equation}
If one will act on arbitrary linear combination of states Eq.\ref{eq:bazis} by the operator
\begin{equation}
\hat{E}=\int{d\mathbf{r}}{{\hat{\bar{q}}}^{+}}\left( \xi ,\mathbf{r} \right){{\hat{q}}^{-}}\left( \xi ,\mathbf{r} \right),
\label{eq:edinichnij}	
\end{equation}
one may note that on the subspace of states which contain one quark and one antiquark the operator in Eq.\ref{eq:edinichnij} acts as unity operator. The same holds for the operator
\begin{equation}
{\hat{E}}'=\int{d\mathbf{r}}{{\hat{q}}^{+}}\left( \xi ,\mathbf{r} \right){{\hat{\bar{q}}}^{-}}\left( \xi ,\mathbf{r} \right).
\label{eq:edinichnij1}	
\end{equation}
which acts on the same subspace. If the first summand in the one-particle part ${{\hat{H}}^{\left( 0 \right)}}$ of the Hamiltonian Eq.\ref{eq:H} is multiplied by the unity operator Eq.\ref{eq:edinichnij1}, and the second summand is multiplied by Eq.\ref{eq:edinichnij} we achieve the expression of the one-particle part in the two-particle effective form
\begin{equation}
\begin{split}
& {{{\hat{H}}}^{\left( 0 \right)}}=\left( -\frac{1}{2m} \right)\int{d{{{\mathbf{r}}}_{1}}d{{{\mathbf{r}}}_{2}}}\left( {{{\hat{\bar{q}}}}^{+}}\left( {{\xi }_{1}},{{{\mathbf{r}}}_{1}} \right){{{\hat{q}}}^{+}}\left( {{\xi }_{2}},{{{\mathbf{r}}}_{2}} \right){{\Delta }_{1}}{{{\hat{\bar{q}}}}^{-}}\left( {{\xi }_{2}},{{{\mathbf{r}}}_{2}} \right){{{\hat{q}}}^{-}}\left( {{\xi }_{1}},{{{\mathbf{r}}}_{1}} \right)+ \right. \\
& \left. +{{{\hat{\bar{q}}}}^{+}}\left( {{\xi }_{1}},{{{\mathbf{r}}}_{1}} \right){{{\hat{q}}}^{+}}\left( {{\xi }_{2}},{{{\mathbf{r}}}_{2}} \right){{\Delta }_{2}}{{{\hat{\bar{q}}}}^{-}}\left( {{\xi }_{2}},{{{\mathbf{r}}}_{2}} \right){{{\hat{q}}}^{-}}\left( {{\xi }_{1}},{{{\mathbf{r}}}_{1}} \right) \right).\\
\end{split}
\label{eq:odnodvux}	
\end{equation}

Next, let us replace the one-particle part of the Hamiltonian Eq.\ref{eq:H} by Eq.\ref{eq:odnodvux} and, after this replacement, pass to the variables Eq.\ref{eq:Jacobi}. We also introduce the following notations
\begin{equation}
\begin{split}
& {{{\mathbf{r}}}_{1}}\left( \mathbf{R},\mathbf{r} \right)=\mathbf{R}-\frac{1}{2}\mathbf{r},~~~{{{\mathbf{r}}}_{2}}\left( \mathbf{R},\mathbf{r} \right)=\mathbf{R}+\frac{1}{2}\mathbf{r}, \\
& {{{\hat{\bar{q}}}}^{+}}\left( {{\xi }_{1}},{{{\mathbf{r}}}_{1}}\left( \mathbf{R},\mathbf{r} \right) \right)={{{\hat{\bar{q}}}}^{+}_{1}},~~~{{{\hat{q}}}^{+}}\left( {{\xi }_{2}},{{{\mathbf{r}}}_{2}}\left( \mathbf{R},\mathbf{r} \right) \right)={{{\hat{q}}}^{+}_{2}}, \\
& {{{\hat{q}}}^{-}}\left( {{\xi }_{1}},{{{\mathbf{r}}}_{1}}\left( \mathbf{R},\mathbf{r} \right) \right)={{{\hat{q}}}^{-}}_{1}. 
~~~{{{\hat{\bar{q}}}}^{-}}\left( {{\xi }_{2}},{{{\mathbf{r}}}_{2}}\left( \mathbf{R},\mathbf{r} \right) \right)={{{\hat{\bar{q}}}}^{-}_{2}}.\\
 \end{split}
\label{eq:pozn27_06}	
\end{equation}

Then, instead of Hamiltonian Eq.\ref{eq:H} we get
\begin{equation}
\begin{split}
& \hat{H}={{{\hat{H}}}^{\left( {\mathbf{R}} \right)}}+{{{\hat{H}}}^{\left( \mathbf{r},V \right)}}, \\ 
& {{{\hat{H}}}^{\left( {\mathbf{R}} \right)}}=\left( -\frac{1}{4m} \right)\int{d\mathbf{R}d\mathbf{r}}{{{\hat{\bar{q}}}}^{+}_{1}}{{{\hat{q}}}^{+}_{2}}{{\Delta }_{{\mathbf{R}}}}{{{\hat{\bar{q}}}}^{-}_{2}}{{{\hat{q}}}^{-}_{1}}, \\ 
& {{{\hat{H}}}^{\left( \mathbf{r},V \right)}}=\int{d\mathbf{R}d\pmb{r}}{{{\hat{\bar{q}}}}^{+}_{1}}{{{\hat{q}}}^{+}}_{2}\left( -\frac{1}{m}{{\Delta }_{{\pmb{r}}}}+V\left( {\pmb{r}} \right) \right){{{\hat{\bar{q}}}}^{-}_{2}} {{{\hat{q}}}^{-}}_{1} .\\
\end{split}
\label{eq:Hvnut_vnech}
\end{equation}

The operator ${\hat{H}}^{\left( {\mathbf{R}} \right)}$ we define as the center of mass Hamiltonian, and the operator ${{\hat{H}}^{\left( \mathbf{r},V \right)}}$ we define as the internal Hamiltonian of the system. Thus the Hamiltonian Eq.\ref{eq:Hvnut_vnech} can be written as

\begin{eqnarray}
\hat{H}=\int{{{{\hat{T}}}_{00}}\left( {\mathbf{R}} \right)d\mathbf{R}},
\label{eq:oprT00} 
\end{eqnarray}
where the energy density operator ${{\hat{T}}_{00}}\left( {\mathbf{R}} \right)$ can be written in the form:
\begin{equation}
{{\hat{T}}_{00}}\left( {\mathbf{R}} \right)=T_{00}^{\left( {\mathbf{R}} \right)}\left( {\mathbf{R}} \right)+T_{00}^{\left( {\mathbf{r}} \right)}\left( {\mathbf{R}} \right)+T_{00}^{\left( V \right)}\left( {\mathbf{R}} \right),
\label{eq:T00RrV}
\end{equation}
with the help of the following denotations:
\begin{equation}
\begin{split}
& T_{00}^{\left( {\mathbf{R}} \right)}\left( {\mathbf{R}} \right)=\left( -\frac{1}{4m} \right)\int{d\mathbf{r}}{{{\hat{\bar{q}}}}^{+}_{1}}{{{\hat{q}}}^{+}_{2}}{{\Delta }_{{\mathbf{R}}}}{{{\hat{\bar{q}}}}^{-}_{2}}{{{\hat{q}}}^{-}_{1}}, \\ 
& T_{00}^{\left( {\mathbf{r}} \right)}\left( {\mathbf{R}} \right)=\left( -\frac{1}{m} \right)\int{d\mathbf{r}}{{{\hat{\bar{q}}}}^{+}_{1}}{{{\hat{q}}}^{+}_{2}}{{\Delta }_{{\vec{r}}}}{{{\hat{\bar{q}}}}^{-}_{2}}{{{\hat{q}}}^{-}_{1}}, \\ 
& T_{00}^{\left( V \right)}\left( {\mathbf{R}} \right)=\int{d\mathbf{r}}{{{\hat{\bar{q}}}}^{+}_{1}}{{{\hat{q}}}^{+}_{2}}V\left( {\mathbf{r}} \right){{{\hat{\bar{q}}}}^{-}_{2}}{{{\hat{q}}}^{-}_{1}}.
\label{eq:T00RrVoboznacenija}
\end{split}
\end{equation}

Relations in Eqs.\ref{eq:oprT00}-\ref{eq:T00RrVoboznacenija} define the nonrelativistic approximation for the operator ${{\hat{T}}_{00}}\left( {\mathbf{r}} \right)$ in the representation of second quantization on Fock subspace.
That approximation can be used to construct nonrelativistic approximation for generator Eq.\ref{eq:M03_ot_T00}. Using these expressions, we have:
\begin{equation}
{{\hat{M}}_{03}}\left( {{{\hat{T}}}_{00}} \right)=\hat{M}_{03}^{\left( {\mathbf{R}} \right)}+\hat{M}_{03}^{\left( {\mathbf{r}} \right)}+\hat{M}_{03}^{\left( V \right)},
\label{eq:oprM03}
\end{equation}
where
\begin{equation}
\hat{M}_{03}^{\left( a \right)}=\int{d\mathbf{R}}\left( {{R}_{3}}T_{00}^{\left( a \right)}\left( {\mathbf{R}} \right) \right),
\label{eq:M03a}
\end{equation}
and index $a$ takes three possible values $a=\mathbf{R},\mathbf{r},V$.

Having a nonrelativistic approximation for the generator Eq.\ref{eq:vt_kv_generator} we can act by the associated operator 
Eq.\ref{eq:generator} on the nonrelativistic approximation for state Eq.\ref{eq:coord_cast} and obtain the probability amplitude of this state in the new reference frame. 

Further we can show that the operators ${{\hat{M}}_{03}}\left( {{{\hat{T}}}_{00}} \right)$ and ${{\hat{H}}^{\left( \mathbf{r},V \right)}}$ commute. A detailed proof of this statement is given in\cite{Deliyergiyev:2013xha}.

We are interested in the ground state of the bound system of quarks, that in center of mass system of quarks corresponds to the eigenvalue which equals the mass of hadron. However, for the aforementioned ground state this eigenvalue is not degenerate. 
Therefore, since that the generator Eq.\ref{eq:vt_kv_generator} commutes with the internal Hamiltonian, we get that the state Eq.\ref{eq:coord_cast} is the eigenstate for both the boost generator Eq.\ref{eq:vt_kv_generator} and the operator Eq.\ref{eq:generator}:
\begin{equation}
{{\hat{M}}_{03}}\left| \mu  \right\rangle ={{m}_{03}}\left| \mu  \right\rangle,
\label{eq:eigenval}
\end{equation}
where ${{m}_{03}}$ is the eigenvalue of generator ${{\hat{M}}_{03}}$ that correspond to the state $\left| \mu  \right\rangle$.

In order to determine the eigenvalue ${{m}_{03}}$, we make use of symmetry features of the eigenstate $\left| \mu  \right\rangle$. 
That is, in particular, this state should transform into itself upon an arbitrary inversion of coordinate exes. Besides, if we suppose that the potential of the quarks and antiquarks interaction is spherically symmetric, then the ground state of this system should also be spherically symmetric, i.e. the state that turn into itself upon an arbitrary rotation. If we denote the unitary operator that represents an inversion or a rotation on the Fock subspace by as ${{\hat{U}}^{\left( I,R \right)}}$, then we have

\begin{equation}
{{\hat{U}}^{\left( I,R \right)}}\left| \mu  \right\rangle =\left| \mu  \right\rangle .
\label{eq:UIR}
\end{equation}
Hence, Eq.\ref{eq:eigenval} can be written as
\begin{equation}
{{\hat{M}}_{03}}{{\hat{U}}^{\left( I,R \right)}}\left| \mu  \right\rangle ={{m}_{03}}{{\hat{U}}^{\left( I,R \right)}}\left| \mu  \right\rangle,
\label{eq:eigenval1}
\end{equation}
or 
\begin{equation}
{{\left( {{{\hat{U}}}^{\left( I,R \right)}} \right)}^{-1}}{{\hat{M}}_{03}}{{\hat{U}}^{\left( I,R \right)}}\left| \mu  \right\rangle ={{m}_{03}}\left| \mu  \right\rangle.
\label{eq:eigenval-1}
\end{equation} 
Operator $\left( \hat{U}^{\left( I,R \right)} \right)^{-1}{{\hat{M}}_{03}}{{\hat{U}}^{\left( I,R \right)}}$ associated with ${\hat{M}}_{03}$ by the tensor transformation rules. It means that we choose an inverse or a rotation, that changes the orientation of $z$ axis to the opposite one, then, we have
\begin{equation}
 {{\left( {{{\hat{U}}}^{\left( I,R \right)}} \right)}^{-1}}{{\hat{M}}_{03}}{{\hat{U}}^{\left( I,R \right)}}=-{{\hat{M}}_{03}}.
\label{eq:minusM03}
\end{equation}

But then, inserting Eq.\ref{eq:minusM03} into Eq.\ref{eq:eigenval-1} with account of Eq.\ref{eq:eigenval}, we get 
\begin{equation}
{{m}_{03}}=0.
\label{eq:ravnonulu}
\end{equation}

So, if we note that $\left| {{\mu }'} \right\rangle$ is the bound state of two-particle system in the reference frame, which is obtained from the c.m.s. of these particles by the boost transformation along $z$ axis with rapidity $Y$, then we have
\begin{equation}
\left| {{\mu }'} \right\rangle =\exp \left( i{{{\hat{M}}}_{03}}Y \right)\left| \mu  \right\rangle. 
\label{eq:mustrih}
\end{equation}
Given Eq.\ref{eq:eigenval}, \ref{eq:ravnonulu}, we see that from the whole series, that represents the result of acting by the operator's exponent $\exp \left( i{{{\hat{M}}}_{03}}Y \right)$ on the state $\left|\mu\right\rangle$, there is the unity-operator summand only, which yields a non-zero result. So, we have
\begin{equation}
\left| {{\mu }'} \right\rangle =\left| \mu  \right\rangle. 
\label{eq:mustrihravnomu}
\end{equation}
This result matches with the result shown in Eq.\ref{eq:visnovok}, which was obtained in the representation of the differential operators. Hence, we draw the conclusion, that the internal state of nonrelativistic system of bound particles does not change
upon the boost transformation with boost to the reference frame, in which this bound system holds a relativistic energy-momentum.

\section {The group-theory approach to the state transformation problem}

In the previous sections we have considered two different representations of the boost generator ${{\hat{M}}_{03}}$ which have led us to the identical results. The question therefore arises whether these results can be generalized. This generalization can be achieved 
if we review the problem using the general group-theoretical concerns. 

Let us consider the generators of Poincare group. We have four generators of space-time translations ${{\hat{P}}_{a}}$, where $a=0,1,2,3$ and six Lorentz generators ${{\hat{M}}_{ab}}=-{{\hat{M}}_{ba}}$. Commutation relations between these generators depend only on the group multiplication law, whereas the above mentioned state-transformation features of quantum systems of interacting particles do not affect these commutation relations. In addition, these commutation relations do not depend on the exact choice of the representation of these generators, so we can examine them without using the explicit form of these generators. 
As it is known from \cite{Raider} the operator ${{g}^{ab}}{{\hat{P}}_{a}}{{\hat{P}}_{b}}$ commutes with all generators of the Poincare group and particularly with the generator ${{\hat{M}}_{03}}$ that we are interested in. 
Take into account also that in accordance with the field quantization postulate \cite{Bogolubov} the generator ${{\hat{P}}_{0}}$ should match with the total Hamiltonian of the system, and the operators ${{\hat{P}}_{b}}$, where $b=1,2,3$, should match with the operators of the momentum components. Then, one may note that the operator ${{g}^{ab}}{{\hat{P}}_{a}}{{\hat{P}}_{b}}$ matches with the square of the internal Hamiltonian of the system, since all eigenvalues of this operator are equal to the corresponding squared eigenvalues of the internal energy of the considered particle system.

As a result, the boost generator commutes with the square of the internal Hamiltonian irrespective of the possibility to apply the nonrelativistic approximation in the c.m.s of the examined particle system. Therefore, it can be argued that any non-degenerate eigenstate of the internal Hamiltonian should turn into itself upon the boost transformation.



\section{Discussion of Results and Conclusions}

It is known that the ground state of the system with the central interaction is spherically symmetric. From the obtained result we conclude that this state will not change during the transition to the new reference frame. In other words, the state remains spherically symmetric and does not undergo to the Lorentz contraction.

Thus the Lorentz contraction in the bound quantum systems does not exist. This conclusion can be very important, since the description of the elastic scattering of hadron-hadron processes rely on the geometric models of discs \cite{Dremin:2013}, which are based on the assumption that the contraction takes place.

The physical reasoning for this follows from the previous discussion. Indeed, when we consider the problem of the rod contraction in the rod's rest frame it is not sufficient whether the coordinates of its ends are measured simultaneously or not. Rather, it is sufficient only to ensure the simultaneity of measurements of the rod's ends coordinates in the reference frame relative to which it is moving. Therefore, the problem of rod contraction can be considered in terms of the relations between the coordinates and time of the same events, measured in the two different reference frames. This is not the case for the problem considered in the paper. In measuring the coordinates of interacting particles in an arbitrary reference frame we have to ensure that the measurements are simultaneous with respect to the reference frame, relative to which the measurements are taken. This makes an essential difference to the problem of rod contraction.
And unlike the rod case, there is no such reference frame, where the measurements of particle coordinates could be taken not at the same time. 
This leads to the fact that we should talk about different measurement events held by different observers. Then the coordinates and time of those events are not connected via Lorentz transformation because these transformations connect the coordinates and time of the same event measured in the different reference frames. Taking into account that via the Lorentz contraction is the exact result of the Lorentz transformation formulas, which do not hold in our case, there is no wonder that we have obtained the result that such a contraction does not take place in the considered problem.

\section{Appendix: The problem of transformation the dependence of length on time.}
\label {AppendixA_reply_part01}

From the argumentations made above it is clear that the classical result of the Lorentz contraction may be altered significantly if one will consider the situation where the rod's length can vary with time, contrary to the problem of the constant-length rod. In this case, in contrast to the ``general" problem of the Lorentz contraction, there is no reference frame with respect to which the coordinates of the rod's ends could be measured at the various time points. In this appendix we would like to demonstrate that the possibility exists that all inertial observers will measure the same dependence of the rod length on time. Thus, we want to show that our result is not specific to a quantum system only and takes place also in the ``regular" special theory of relativity. 

However, it will be more convenient if instead of examining the solid-rod problem, we turn to the problem of the distance between two particles, motion of which is governed by the given law. 

Namely, suppose that there is particle in some inertial reference frame which moves along the $x$-axis according to the given law $x(t)$. And suppose, that there is another particle with same mass in that reference frame, which moves according to the given law $(-x(t))$. Thus, the considered reference frame in the following example would be the center of mass system for these particles. In this frame of reference the distance between those particles at the time, $t$, would be $l(t)=2x(t)$. Now, let's try to find the dependence of the distance between particles as function of time $t$ which will be measured by the inertial observer who uses as a frame of reference the one that is boosted along the $x$-axis with respect to the original frame with velocity $v$.


In order to compute the dependence $l^{\prime}(t^{\prime})$ in the other frame we will fix the time moment, $t^{\prime}$, in which the measurement of the particle's coordinates will be done in the original reference frame. 
Then, the measurement of coordinate of the first particle by the clock of the original reference frame will be done in the moment $t_{1}(t^{\prime})$, which is expressed through $t^{\prime}$ by the following equation:
\begin{equation}
t^{\prime}=\frac{t-\frac{x(t)v}{c^{2}}}{\sqrt{1-\frac{v^2}{c^2}}}
\label{eq:appendixA_01}
\end{equation}
Similarly, for the second particle measuring its coordinates at time, $t^{\prime}$ will corresponds to time,$t_{2}(t^{\prime})$, that can be obtained from solution of the following equation:
\begin{equation}
t^{\prime}=\frac{t+\frac{x(t)v}{c^{2}}}{\sqrt{1-\frac{v^2}{c^2}}}
\label{eq:appendixA_02}
\end{equation}
Then, the coordinates of the particles, $x_{1}^{\prime}(t^{\prime})$ and $x_{2}^{\prime}(t^{\prime})$, measured simultaneously using the clock from the ``primed" reference frame can be obtained as follows:
\begin{equation}
x_{1}^{\prime}(t^{\prime})=\frac{x_{1}\left( t_{1}(t^{\prime})\right) - vt_{1}(t^{\prime})}{\sqrt{1-\frac{v^2}{c^2}}},~~~ x_{2}^{\prime}(t^{\prime})=\frac{x_{2}\left( t_{2}(t^{\prime})\right) - vt_{2}(t^{\prime})}{\sqrt{1-\frac{v^2}{c^2}}}
\label{eq:appendixA_03}
\end{equation}
Calculating the absolute value of the difference of these coordinates we obtain the sought dependence, $l^{\prime}(t^{\prime})$. Herewith, similar to the problem of transformation of the quantum-mechanical state, in which one examines the dependences $l(t)$ and $l^{\prime}(t^{\prime})$, we can not establish any relations between the arguments $t$ and $t^{\prime}$, because in this case, as well as in the quantum mechanical measurement of particles' coordinates one may use only different events.  Events that are simultaneous in one reference frame will not be simultaneous in another, therefore, these events are not suitable for the measurement. At the same time, in the original problem of the rod contraction, indeed one can use the same events because in the rod's rest frame the measurements can be indeed performed not simultaneously.

Taking the above mentioned concerns into account, the only entity one may compare is the form of functional dependence of the distance between the particles with respect to time measured in the different reference frames. It is similar to the problem of comparing the form of functional dependence of the probability amplitude with respect to its arguments in the different inertial frames.

We now consider the special case of the above reasoning, when
\begin{equation}
x(t)=\eta c t_{0} \sqrt{1+\left( \frac{t}{t_{0}} \right)^{2}},~~ t \in [-\infty, +\infty],~~ 0\textless \eta \leq 1 
\label{eq:appendixA_04}
\end{equation}
Such dependence is chosen for reasons of simplicity of solutions of emerging equations. 
For the velocity of the particle in the considered reference frame we have the relation $\vert \frac{dx(t)}{dt} \vert \leq \eta c$, i.e. the velocity at any given time does not exceed $c$, even if $\eta=1$.

Nondimensionalizing the space coordinate with $ct_{0}$, the time coordinate with $t_{0}$ (where $t=t_{0}\tau$, $\tau$ is dimensionless time) and the velocity with $c$, and by doing the same procedures described above we obtain, instead of Eq.\ref{eq:appendixA_01}, the distance between two particles as a function of time:
\begin{equation}
l^{\prime}(\tau^{\prime})=\frac{2\eta \sqrt{1-v^2}} {1-v^2\eta^2} \sqrt{1-v^{2}\eta^2 + \left( 1 - v^{2} \right){\tau^{\prime}}^{2}}
\label{eq:appendixA_08}
\end{equation}
If $\eta \textless 1$ then for any value of the argument $\tau$ we have $l^{\prime}(\tau)\textless l(\tau)$. However, if in Eq.\ref{eq:appendixA_08} we set $\eta=1$, that as we have seen is acceptable, then we obtain
\begin{equation}
l^{\prime}(\tau^{\prime})=2\sqrt{1+{\tau^{\prime}}^{2}}
\label{eq:appendixA_09}
\end{equation}
Thus, due to the fact that velocity has dropped out of Eq.\ref{eq:appendixA_09}, the dependence of rod's length on time has the same form in all inertial systems, just as in our work, where the probability amplitude has the same dependence on its arguments.

As you may noticed, this result was obtained on the basis of Lorentz transformations only. 
Therefore, this result is equally plausible as one of the rod contraction, being the consequence of the same transformations. And these results do not contradict each other.

\normalsize {\textbf{REFERENCES}}
\bibliography{PhysRev_references}

\end{document}